\documentstyle[multicol,aps,psfig,epsf]{revtex}

\newcommand{\beq}{\begin{equation}}
\newcommand{\eeq}{\end{equation}}
\newcommand{\bdis}{\begin{displaymath}}
\newcommand{\edis}{\end{displaymath}}
\newcommand{\bea}{\begin{eqnarray}}
\newcommand{\eea}{\end{eqnarray}}                 
\newcommand{\barr}{\begin{array}}
\newcommand{\earr}{\end{array}}

\begin{document}
\twocolumn[\hsize\textwidth\columnwidth\hsize\csname 
 @twocolumnfalse\endcsname

\title{Entropy for relaxation dynamics in granular media}
\author{Emanuele Caglioti$^{1}$ and Vittorio Loreto$^{2}$}
\pagestyle{myheadings}
\markboth{Caglioti, Loreto}{Caglioti, Loreto}
\address{
$^1$Dipartimento di Matematica, Universit\'a di 
Roma ``La Sapienza'', Piazzale Aldo Moro 2, 00185 Roma, 
Italy}
\address{
$^2$P.M.M.H. Ecole Sup\'erieure de Physique et Chimie Industrielles,
10, rue Vauquelin, 75231 Paris CEDEX 05 France}
\maketitle
\date{\today}
\maketitle
\begin{abstract}
We investigate the role of entropic concepts for the
relaxation dynamics in granular systems. 
In these systems the existence of a {\em geometrical} 
frustration induces a drastic modification of the allowed phase space,
which in its turn induces a dynamic behavior characterized by 
hierarchical relaxation phenomena with several time scales associated.
In particular we show how, in the framework of a mean-field model 
introduced for the compaction phenomenon, there exists a 
{\em free-energy}-like functional which decreases along the 
trajectories of the dynamics and which allows to 
account for the asymptotic behavior: e.g. density profile, 
segregation phenomena. Also we are able to perform the
continuous limit of the above mentioned model which turns out to be 
a diffusive limit. In this framework one can single out two separate 
physical ingredients: the {\em free-energy}-like functional that defines 
the phase-space and the asymptotic states and a diffusion coefficient 
$D(\rho)$ accounting for the velocity of approach to the asymptotic 
stationary states.
\end{abstract}
\smallskip
\vskip2pc]


Granular media enter only partially in the framework
of equilibrium statistical mechanics and hydrodynamics. 
Their dynamics constitutes a very complex problem 
of non-equilibrium which poses novel questions and challenges to 
theorists and experimentalists\cite{grani,Mehta}.

Generally speaking granular materials cannot be described as 
equilibrium systems neither from the configurational point of 
view nor from the dynamical point of view.
It is known in fact that these systems remain easily trapped
in some metastable configurations which can last for long
time intervals unless they are shaken or perturbed. 
A granular system may be in a number of different microscopic 
states at fixed macroscopic densities, and, more in general, for 
a given ensemble of macroscopic parameters. Many unusual properties 
are linked to this non trivial packing \cite{grani}.
The configurational space of these systems is very complex and presents
a structure with several local minima. This structure induces 
a dynamic behavior characterized by hierarchical relaxation phenomena
with several associated time scales.
A general mechanism bringing to the existence of such a structure is
based on the concept of {\em frustration}, that for instance in 
granular media has a geometrical origin. The existence of 
complex geometrical interactions between the grains induces a rough 
landscape in the structure of the allowable phase
space and in the configurational entropy. 
In their turn these effects induce the need of complex
cooperative rearrangements which account for the very 
slow relaxation dynamics of these systems.   
At high densities (or very low temperatures for thermal systems)
the system remains trapped in a local minimum and exhibits
a non-ergodic behavior as well as very slow relaxations:
the logarithmic compaction in granular media 
\cite{Knight,NCH,Ben-Naim,prl,DeGennes} 
or the Kohlrausch-Williams-Watts (KWW) relaxations in glassy
systems\cite{vetri}.

In this paper we try to elucidate the role that the concept of 
entropy can play in the dynamics of these systems. 
In particular we show how, in the framework of a 
mean-field model introduced for the compaction phenomenon, 
there exists a {\em free-energy}-like functional which decreases along the 
trajectories of the dynamics and which allows to 
account for the asymptotic behavior: e.g. density profile, 
segregation phenomena. Furthermore the continuous limit 
of the above mentioned models allows us to comment on the 
relationship between entropic and dynamical effects in explaining 
the relaxation phenomena in granular media \cite{Mehta,Edwards,Hans}.

 
We consider a simple model which describes the evolution of a system 
of particles which hop on a lattice of $k=0,...,N$ stacked planes, 
as introduced in \cite{prl}.
In particular the system represents an ensemble of particles which can move up 
or down in a system of $N$ layers in such a way that their total number 
is conserved. We ignore the correlations among particles
rearrangements and problems related to the mechanical stability of
the system. 
The master equation for the density on a generic plane $k$, except for
the $k=0$ plane, is given by:
\begin{eqnarray}
\partial_t \rho_k &=& 
(1-\rho_k)D(\rho_k) [ \rho_{k-1} \cdot p_{up} +\rho_{k+1} \cdot
p_{down}) ]+\nonumber \\
&& -\rho_k [(1-\rho_{k-1}) D(\rho_{k-1}) p_{down} +\nonumber\\ 
&& +(1-\rho_{k+1}) D(\rho_{k+1})p_{up}] 
\label{n-plan}
\end{eqnarray}
where $p_{down}$ and $p_{up}$ (with $p_{up}+p_{down}=1$)
represents the probability for the particles to move downwards
or upwards, respectively, among the different planes.
With $p_{up}$ and $p_{down}$ we can define the quantity 
$x=p_{up}/p_{down}$ which quantifies the importance of gravity
in the system. We can also associate to $x$ a sort of temperature for 
the system given by $T \sim 1 / log(1/x)$. We shall return to this 
point later on.

$D(\rho_{k})$  represents a sort of mobility
for the particles given by the probability that the particle could
find enough space to move.  Apart from other effects it mainly 
takes onto account the geometrical effects of frustration, i.e. 
the fact that the packing prevents the free movement of 
the particles. 

Later on we shall show how the analysis reported here is very general and 
does not depend on the exact functional form chosen for  $D(\rho)$.
For the sake of clarity before discussing the problem in its generality we shall 
consider a possible functional form for $D(\rho_{k})$ suitable for the 
compaction problem.
In a naive way one could imagine a functional form 
like $D(\rho_{k})= \rho_k (1- \rho_{k^{\prime}})$ 
obtained by considering only the interactions with the nearest 
neighbors planes whose density is designed by $\rho_{k^{\prime}}$.
It is easy to realize that such an approach
does not account for the complexity of the problem where the packing 
at high densities creates long range correlations in the system,
and, using this functional form, the equations show a trivial relaxation. 
A possible general form of $D(\rho_{k})$, which can be seen as the
outcome of a theory based on the existence of regions of cooperativity
\cite{nostroadam} (as well as of a free-volume theory
\cite{Ben-Naim,DeGennes}) for granular media, includes a term like
\beq
D(\rho_{k})=D_0\exp[-\alpha/(1-\rho_{k})].
\label{drho}
\eeq
The parameter $\alpha$ quantifies how much the shape and the
dimensions of the particle frustrate its motion. 
Higher is $\alpha$ higher will be the geometrical frustration felt by
the corresponding particle. For instance in the problems of parking of 
$r$-mers (i.e. segments of length $r$) 
on a line, under the hypothesis of an exponential distribution
for the lengths of the empty (or filled) intervals, 
the probability for each $r$-mer to find a
sufficient space to land is $\exp(-{r}/ {(1-\rho}))$, where $\rho$ 
is the occupation density on the line \cite{Ben-Naim}.

The question we want to address is whether there exists a variational
principle driving the relaxation phenomena in this system and in
general in granular media.
In other words one could ask if, in analogy with what happens 
for a liquid system, there exists some free-energy-like functional 
minimized (Lyapunov functional) \cite{Lyapunov} by the dynamical evolution.
In a very general way it is possible to write explicitly
a Lyapunov functional for the system of equations (\ref{n-plan}), 
say the functional which decreases 
monotonically along the trajectories of the motion.
This functional can be cast in the form of a free-energy-like function:
\begin{equation}
F=\sum_{k=0}^{\infty} [ \gamma(x) k \rho_k - S(\rho_k) ]
\label{functional}
\end{equation}
where the $S(\rho_k)$, the {\em entropic}-like contribution,
and $\gamma(x)$ have to be determined in a self-consistent way
by imposing that $F$ decreases for any exchange 
of particles between two generic planes $k$ and $k+1$.
Writing explicitly the expression for $dF$ for a generic particle 
exchange and imposing $dF \le 0$,  
for the particular choice of $D(\rho_{k})$ we made,
after some manipulations one gets for the functional (\ref{functional}) the expressions
\begin{eqnarray}
S(\rho_k) =\rho_k log\big(\frac{1-\rho_k}{\rho_k}\big), \,\,\,\,\,\,\,
\gamma(x)  =   log(\frac{1}{x}).
\label{funcmono}
\end{eqnarray}
Let us stress how $S(\rho_k)$ has exactly  the same (critical) behavior,
as $\rho_k \rightarrow 1$,  of the entropy of a one-dimensional continuous 
system, filled with bars of unitary length at a certain density $\rho$.
The functional $F$ is a concave function because 
$\partial^2 F / \partial \rho_k^2 \ge 0 \,\,\, \forall k$
\cite{hessian}.

As a consequence there exists a unique minimum for this
functional and this minimum should correspond to the stationary
state of the system. In the case of a monodisperse system
we expect that the stationary state (i.e. the density profile for the
system) is the one obtained, in a continuous process of shaking, 
after a very long transient.

In the general case with arbitrary $N$, and for $N \rightarrow +\infty$, 
it is possible to get the exact asymptotic stationary solution 
for the density on each plane. We denote with $M$ the 
total ``mass'' of the system, i.e. the maximal number of planes which can be 
completely filled.
Using a standard Lagrange multiplier method,
where one tries to find the extremum of $F$ subject to
to the constraint $\sum_k \rho_k =M$, the solution is given 
by the following implicit expression:
\beq
f(\rho_k) = f(\rho_0) x^k
\eeq
where 
\beq
f(s)= \frac{s}{1-s} e^{\frac{1}{1-s}},
\eeq
and $\rho_0$ is the density on the zero-th plane which is a complex
function of the total mass of the system.
In order to visualize the solution it is possible to extract 
the approximate explicit behaviors. 
In particular one gets:
\begin{eqnarray}
\rho_k^{\infty} & \simeq & 1 - 1/[(M-k) \cdot \log(1/x)] \,\, \mbox{for} \,\,
k << M\\\nonumber\\
\rho_k^{\infty} & \simeq & e^{(M-k) \cdot \log(1/x)}\,\, \mbox{for} \,\,
k >> M.
\label{stationary}
\end{eqnarray}

The stationary solution tends thus to a step function $\theta(k-M)$ in
the limit $x \rightarrow 0$.
This behavior is very well verified in the experiments\cite{clement}.

Let us now ask what happens considering a bidisperse 
system, i.e. by considering two kind of particles 
defined by two different values of $\alpha$
($\alpha_s=1,\alpha_b=\alpha=2$ in the
specific case).
In the case the master equations for the densities
$\rho_k^b$ and $\rho_k^s$ on a generic plane $k$, except for the plane
$k=0$, are given by:
\begin{eqnarray}
\partial_t \rho_k^{b,s} &=& 
D^{b,s}(\rho_k) [ \rho_{k-1}^{b,s} p_{up} +\rho_{k+1}^{b,s} p_{down})
]+ \nonumber \\ && 
-\rho_k^{b,s} [D^{b,s}(\rho_{k-1}) p_{down} + D^{b,s}(\rho_{k+1}) p_{up}]
\label{dymo}
\end{eqnarray}
where $\rho_k=\rho_k^s + \alpha \rho_k^b$ 
is the total density on the $k$-th plane and 
$D^{b,s}(\rho_k)$ are the probabilities for a
particle $s$ or $b$ respectively, landing on the plane $k$, 
to fit the local geometrical environment. 
In this sense $D^{b,s}(\rho)$ are the analog of the mobilities
for the two kind of particles and they take into account
the cooperative effects on the dynamics generated by the frustration.
A possible functional form for $D^{b,s}(\rho)$ 
(see for a similar approach \cite{Ben-Naim,prl,DeGennes,nostroadam}), 
is given by $D^s(\rho) \sim (1-\rho) \exp[- 1/(1-\rho)]$ and
$D^b(\rho) \sim (1-\rho)^{\alpha} 
\exp[- {\alpha}/(1-\rho)]$ with $\alpha \ge 1$.

Let us now look for a Lyapunov functional equivalent to
Eq.(\ref{functional}) in this polydisperse case which now will
have the form:
\begin{equation}
F=\sum_{k=0}^{\infty} [ \gamma(x) n \rho_k - S(\rho_k^b,\rho_k^s) ]
\label{functional2}
\end{equation}
where the $S(\rho_k^b,\rho_k^s)$ is again an {\em entropic}-like
contribution.
By repeating the calculation along the same lines as before
one gets
\begin{eqnarray}
S(\rho_k^b,\rho_k^s) &=&- \rho_k^s (log(\rho_k^s) -1)
- \rho_k^b (log(\rho_k^b) -1)\nonumber \\
&& -(1-\rho_k) (log(1-\rho_k)-1)+log(1-\rho_k)
\label{entropy}
\end{eqnarray}
and $\gamma(x)=log(1/x)$.

\begin{figure}[h]
\centerline{\psfig{figure=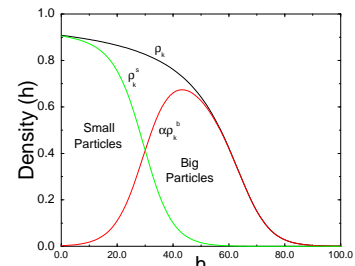,width=5cm,angle=-90}}
\vspace{0.cm}
\caption{Asymptotic density profile for the two species of particles
  in the $N$-planes model described in the text. It is clearly visible
the effect of segregation of the more frustrated (big) particles $\rho_b^k$
on top of the less frustrated (small) ones $\rho_b^k$.} 
\label{segfig}
\end{figure}

Even in this case the functional $F$ is concave and there exists
a unique solution corresponding to the asymptotic stationary state
for the system.
Using a standard Lagrange multiplier method one gets the result:

\begin{eqnarray}
\rho_k^b &=& \rho_k \frac{h_0 (1-\rho_k)^{\alpha-1} 
e^{-\frac{\alpha-1} {1-\rho_k}}} 
{1+\alpha h_0 (1-\rho_k)^{\alpha-1} 
e^{-\frac{\alpha-1} {1-\rho_k}}}\nonumber\\
\rho_k^s & = & \rho_k - \alpha \rho_k^b
\label{solution}
\end{eqnarray}
where  $h_0=h_0(\rho_0^b,\rho_0^s)$ is a constant which can be
estimated by using the total mass $M$ of the system:
\begin{equation}
h_0(\rho_0^b,\rho_0^s)=(1-\rho_0)^{1-\alpha}
e^{\frac{\alpha-1}{1-\rho_0}} \frac{\rho_0^b}{\rho_0^s}.
\end{equation}
$\rho_k$ is given by a non trivial function of $k$ that can be
obtained exactly but in a implicit form. 
$\rho_k$ is a monotonically decreasing function with the
maximum for $k=0$. Fig.~\ref{segfig} shows an example of the
asymptotic stationary solution obtained  from (\ref{solution}) 
with $h_0\simeq 10^3$ corresponding to $x\simeq 0.82$, 
$\rho_0 = 1- \frac{c}{M log(1/x)} \simeq 0.9$ and 
$\frac{\rho_0^b}{\rho_0^s} \simeq 1.5 \cdot 10^{-3}$.
In particular it shows $\rho_k^b$ and $\rho_k^s$ as a function
of $\rho_k$. In this way the effect of segregation
\cite{segtet} of the ''big'' particles (with $\alpha=2$) on top of 
the ''small'' ones (with $\alpha=1$) is clearly visible.
The introduction of two different weights for the two species, say
$p_{up}^{b,s}$ and $p_{down}^{b,s}$, gives rise to a rich phase space
$(p_{up}^{b},p_{up}^{s},\alpha)$ with different regions corresponding
to different behaviors for segregation.
 
Eq.(\ref{entropy}) makes evident how in the dynamics of granular
media entropic effects are far from being negligible and the global
behavior is given by a complex interplay between gravitational and
entropic-like effects.
 
Let us now continue to push forward this point of view and ask whether the 
asymptotic stationary state corresponds to some sort of equilibrium state.
At the equilibrium, i.e. asymptotically, one would expect that for any 
exchange of particles among different planes $dF=0$ for the free-energy-like
functional (\ref{functional2}), or equivalently $dS/dE=1/T=const.$.
By defining the total energy $E= \sum_{k=0}^{\infty} k \rho_{k}$ and
the total entropy $S_{tot} = \sum_{k=0}^{\infty} S(\rho_{k})$, for any 
exchange of particles of type $b$ or $s$ between the planes $k$ and $k+1$
one has:
\beq
\frac{d S_{tot}^{b,s}}{dE} = \frac{\partial S(\rho_{k+1})}
{\partial  \rho_{k+1}^{b,s}} -
\frac{\partial S(\rho_{k})}{\partial  \rho_{k}^{b,s}}.
\label{temp}
\eeq 
It is easy to realize how in the asymptotic stationary state one has
\beq
\frac{d S_{tot}^{s}}{dE} = \frac{d S_{tot}^{b}}{dE} =  
log(\frac{1}{x}) = const.,
\label{temp2}
\eeq
or, what is the same, $\frac{dS_{tot}^{b,s}}{dE}= const.$ everywhere 
in the system  no matter which kind of particles we use for its 
definition.
It is then tempting to associate $log(\frac{1}{x})$ with  the inverse of a 
temperature for the system. The analogy is made stronger by recalling
that $1/log(\frac{1}{x})$ is the usual quantity associated to a temperature
in Monte-Carlo dynamics. From this point of view the relaxation process 
of this system would correspond to an equilibration procedure in which 
$\frac{d S_{tot}^{b,s}}{dE}$ is made uniform everywhere. 
In the specific case of segregation one is forced to think that the system 
evolves in such a way that asymptotically all the particles have the same 
mobility, measured in this case in terms of the entropic change for every 
given displacement.


Let us now generalize the results obtained so far.
Eq.(\ref{n-plan}) represents a particular case of a general class 
of equations that can be written as:
\begin{eqnarray}
\partial_t \rho_k &=& 
g(\rho_k) [ f(\rho_{k-1}) p_{up} + f(\rho_{k+1}) p_{down})
]+ \nonumber \\ && 
-f(\rho_{k}) [g(\rho_{k-1}) p_{down} + g(\rho_{k+1}) p_{up}]
\label{dymogen}
\end{eqnarray} 
where $f$ and $g$ are generic functions for which we only require
$f \ge 0$, $g \ge 0$, $ d f/  d\rho \ge 0$, $ d g /d \rho \le 0$. 
All the results we have shown in the particular case (\ref{n-plan}) are valid 
under these general assumptions.
In particular we can prove that there does exist a functional
that decreases along the trajectories of the motion and its expression
is given by Eq.(\ref{functional}) with:
\begin{eqnarray}
S(\rho_k) = \int_{\rho_k}  log {g(\rho) \over f(\rho)}  d\rho ,\,\,\,\,\,\,\,\,
\gamma(x)  = log(\frac{1}{x}).
\label{funcmonogen}
\end{eqnarray}


A deeper insight in the above mentioned phenomenology is obtained 
by considering the continuum limit for the model described by 
Eq.(\ref{dymogen}). 
More precisely we consider a diffusive limit that consists in scaling 
the space variable as $1\over \epsilon$, the 
time variable as ${1\over \epsilon^{2}}$ and the drift term 
$p_{down}-p_{up}$ее as $\epsilon$.
Therefore $x=\epsilon k$, $\tau=\epsilon^{2}k/2$, 
$p_{down}-p_{up}=\epsilon{\beta}/2$ее and we consider 
the evolution of $u(x)\equiv\rho(k)$.

We get the continuum limit by taking the Taylor expansion of the right 
member of Eq.(\ref{dymogen}) around $x=k\epsilon$. For example 
$\rho(k+1)\equiv u(x+\epsilon)=u(x)+\epsilon\partial_{x}uе+
{1\over 2}\epsilon^{2}\partial_{xx} u+O(\epsilon^{3})$.
   
We get, formally, 
\beq 
\partial_{\tau}u(x)=\beta\partial_{x}е(f g) + \left(g 
\partial_{xx}f - f  \partial_{xx}еg\right) + O(\epsilon),
\eeq
which, in the limit $\epsilon\rightarrow 0$, gives
\beq
\partial_{\tau}u(x)=\beta\partial_{x}е(f g) +(g 
\partial_{xx}еf -f  \partial_{xx}еg) .
\label{continuum1}
\eeq
This is a non linear diffusion equation that may be conveniently 
written in the following form
\beq 
\partial_{\tau}u=\partial_{x}\left(D(u)ее
\partial_{x}{\partial F\over\partial_{u}}\right),
\label{continuum2}
\eeq
where $D=f g$, ${\partial F\over\partial_{u}}$ denotes the functional 
derivative of $F$ with respect to $u$, 
$F=\int_0^{\infty} (\beta\phantom{,} u\phantom{,}x - S(u)) dx,$е
and $S'=\log\left({g\over f}\right)$.
Notice that the functional $F$ decreases with the dynamics induced by 
Eq.~(\ref{continuum1}). One has, in fact,
\beq 
\partial_{\tau}F=\int dx {\partial 
F\over\partial u }\partial_{\tau}u=\int dx {\partial 
F\over\partial  u }\partial_{x}\left(D(u)ее
\partial_{x}{\partial F\over\partial u }\right)
\label{cahn}
\eeq
that, after an integration by parts, gives
\beq 
-D(u) \left({\partial F\over\partial_{u}}\right)^{2}\leq 0.
\eeq
Therefore there exists a ``free energy''-like  functional, $F$, 
for Eq.~(\ref{continuum2}) which has exactly the same form of the
functional defined for the discrete model (see Eq.(\ref{funcmonogen})).
We can notice that while the functional form of $S$ and the value of 
$\beta$ determine in a unique way the asymptotic state they are not 
sufficient to determine the dynamical behavior of the system.
In particular in order to know it one should know the functional form 
of $D(\rho)$.


What we have discussed so far suggests the possibility of introducing, for 
non-thermal systems as granular media, equilibrium 
concepts as free-energy, entropy and temperature. 
More precisely it is possible (in the case studied here) 
to predict the asymptotic state by means of the 
minimization of a suitable functional which can be constructed by  
entropic arguments.
It is worth stressing how granular systems often exhibit memory and so
the existence of a unique Lyapunov functional is not guaranteed in general.
In general one could expect that several Lyapunov functionals are 
associated to different stationary states reached with different dynamical paths.

Finally let us notice how Eq.(\ref{continuum2}) allows us to comment
on the relationship between entropic and dynamical effects in
explaining the relaxation phenomena in granular media.
Once the Lyapunov functional is known (and thus also the entropic 
properties of the system) it is possible to predict
the asymptotic state. However one cannot specify whether the asymptotic state is reached 
in a finite time unless one knows the connectivity
properties of the phase space, which in our case corresponds 
to know $D(\rho)$. {\large Acknowledgements:} the authors wish to thank 
A. Coniglio, S. Krishnamurthy, H.J. Herrmann, M. Nicodemi and S. Roux 
for interesting discussions. 
VL acknowledges financial support under project ERBFMBICT961220.
This work has also been partially supported from the European
Network-Fractals under contract No. FMRXCT980183.

\begin{thebibliography}{99}


\bibitem{grani} see e.g. H.M. Jaeger and S.R. Nagel, {\em Science} {\bf 255},
  1523 (1992);  H.M. Jaeger, S.R. Nagel and R.P. Behringer, {\em Phys. Today}
  April 1996. 

\bibitem{Mehta} A. Mehta, ed., {\em Granular Matter: an 
interdisciplinary approach}, (Springer-Verlag, New York, 1994).

\bibitem{Knight} J.B. Knight, C.G. Fandrich, C. Ning Lau, H.M. Jaeger,
  S.R. Nagel,  {\em Phys. Rev. E} {\bf 51}, 3957 (1995). 

\bibitem{NCH} M. Nicodemi, A. Coniglio, H.J. Herrmann, 
{\em Phys. Rev. E}, {\bf 5}, 3962 (1997); {\em J. Phys. A} {\bf 30}, 
L379 (1997); {\em Physica A} {\bf 240}, 405 (1997). 

\bibitem{Ben-Naim} E. Ben-Naim, J.B. Knight and E.R. Nowak,
{\em Physica D} {\bf 123}, 380 (1998); P.L Krapivsky and E. Ben-Naim, 
{\em J. of Chem. Phys.} {\bf 100}, 6778 (1994).  

\bibitem{prl} E. Caglioti, V. Loreto, H.J. Herrmann and
  M. Nicodemi, {\em Phys. Rev. Lett.} {\bf 79}, 1575 (1997).

\bibitem{DeGennes} T. Boutreux and P.G. de Gennes, in 
{\em Powders and Grains 97}, edited by R. Behringer and J. Jenkins 
(A.A. Balkema, Rotterdam), pp.439.

\bibitem{vetri} M. M\'ezard, G. Parisi and M.A. Virasoro, {\em
    Spin glass theory and beyond}, World Scientific (Singapore 1987).
A. Coniglio: {\em Clusters and Frustration in
    glass-forming systems and granular materials}, Prooc. of the
  Int. School of Physics ``E. Fermi'', Varenna, July 9-19 (1996).

\bibitem{Edwards} S. F. Edwards, {\em J. Stat. Phys.} {\bf 62} (1991) 889; 
Mehta A. and  Edwards S. F., {\em Physica A} {\bf 157} (1989) 1091.

\bibitem{Hans} H.J. Herrmann, {\em J. Physique II} {\bf 3}, (1993) 427.

\bibitem{nostroadam} E. Caglioti, A. Coniglio, H.J. Herrmann,
  V. Loreto and M. Nicodemi,  {\em Physica A} {\bf 265}, 311 (1999).

\bibitem{Lyapunov} for a general introduction see: D.W. Jordan and P. Smith,
{\em Nonlinear Ordinary Diferential Equations}, Oxford, England, 
Clarendon Press (1977).


\bibitem{hessian} The functional $F$ is concave because is given by the sum of 
a linear term, which is irrelevant for the concavity, and of a concave term $S(\rho_k)$.
Since the sum of concave functions is concave one gets the desired result.
On the other hand  one can look at the hessian matrix and realize that it is 
diagonal with all diagonal terms given by $\partial^2 F / \partial \rho_k^2 = 
\partial^2 S / \partial \rho_k^2 \ge 0 \,\,\, \forall k$.



\bibitem{clement} E. Clement and J. Rajchenbach, {\em Europhys. Lett.}
 {\bf 16}, 133 (1991).

\bibitem{segtet} A. Rosato, K.J. Strandburg, F. Prinz and R.H. Swendsen,
{\em Phys. Rev. Lett.} {\bf 58}, 1038 (1986); E. Caglioti, A. Coniglio, H.J. Herrmann, 
V. Loreto and M. Nicodemi, {\em Europhys. Lett.} {\bf 43}, 591 (1998).

\end{thebibliography}
\end{document}